\documentclass[aps,pra,reprint,citeautoscript,showpacs,floatfix]{revtex4-1}

\usepackage{amssymb}
\usepackage{amsmath}
\usepackage{amsfonts}

\newcommand{\ten}[1]{\mathsf{#1}}
\renewcommand{\vec}[1]{\boldsymbol{#1}}
\newcommand{\vech}[1]{\hat{\vec{#1}}}

\newcommand{\Dp}[1]{\partial_{#1}}

\newcommand{\beq}{\begin{eqnarray}}
\newcommand{\eeq}{\end{eqnarray}}
\renewcommand{\d}{{\rm{d}}}

\newcommand{\half}{\frac{1}{2}}
\newcommand{\aeq}{\approx}

\newcommand{\rcite}[1]{Ref. \onlinecite{#1}}
\newcommand{\Rcite}[1]{Ref. \onlinecite{#1}}

\newcommand{\Rcites}[1]{Refs \onlinecite{#1}}

\newcommand{\xrm}{\rm{x}}
\newcommand{\crm}{\rm{c}}
\newcommand{\xc}{\rm{xc}}

\newcommand{\Ec}{E^{\crm}}
\newcommand{\Ex}{E^{\xrm}}
\newcommand{\Exc}{E^{\xc}}

\newcommand{\fxc}{f^{\xc}}
\newcommand{\tFxc}{\ten{F}^{\xc}}

\newcommand{\pr}{^{\prime}}

\newcommand{\rp}{r\pr}

\newcommand{\vx}{\vec{x}}

\renewcommand{\vr}{\vec{r}}
\newcommand{\vrh}{\vech{r}}
\newcommand{\vrp}{\vec{r}\pr}
\newcommand{\vrhp}{\vech{r}\pr}

\newcommand{\vj}{\vec{j}}

\newcommand{\vnu}{\vec{\nu}}
\newcommand{\nur}{\nu^{r}}
\newcommand{\nup}{\nu^{\perp}}
\newcommand{\Fr}{F^{r}}
\newcommand{\Fp}{F^{\perp}}

\newcommand{\vF}{\vec{F}}

\newcommand{\tP}{\ten{P}}
\newcommand{\tV}{\ten{V}}

\newcommand{\vnabla}{\vec{\nabla}}

\newcommand{\starr}{\star_r}

\newcommand{\KS}{{\rm{KS}}}

\newcommand{\EXX}{{\rm{EXX}}}

\newcommand{\VKS}{V^{\KS}}
\newcommand{\hh}{\hat{h}}
\newcommand{\n}{n^0}

\begin{document}
\title{Correlation energies beyond the random-phase approximation: ISTLS applied to spherical atoms and ions}
\author{Tim Gould}\affiliation{Qld Micro- and Nanotechnology Centre, %
Griffith University, Nathan, Qld 4111, Australia}
\author{John F. Dobson}\affiliation{Qld Micro- and Nanotechnology Centre, %
Griffith University, Nathan, Qld 4111, Australia}
\begin{abstract}
The inhomogeneous Singwi, Tosi, Land
and Sjolander (ISTLS) correlation energy functional of
Dobson, Wang and Gould [PRB {\bf 66} 081108(R) (2002)]
has proved to be excellent at predicting correlation energies in
semi-homogeneous systems, showing promise as a robust `next step'
fifth-rung functional by using dynamic correlation to go beyond
the limitations of the direct random-phase approximation (dRPA),
but with similar numerical scaling with system size.
In this work we test the functional on spherically symmetric,
neutral and charged atomic systems and
find it gives excellent results (within 2mHa/$e^-$ except Be)
for the absolute correlation energies of the neutral atoms tested,
and good results for the ions (within 4mHa/$e^-$ except B${}^+$).
In all cases it performs better than the dRPA. When combined with
the previous successes, these new results point to the ISTLS functional
being a prime contender for high-accuracy, benchmark DFT correlation
energy calculations.
\end{abstract}
\maketitle

Since their development,
density-functional theory\cite{HohenbergKohn,KohnSham} (DFT)
methods have vastly increased the range of quantum
mechanical problems that can be studied. This wide range comes
through the use of approximation to the exchange-correlation (xc)
physics necessarily introduced to make Kohn-Sham (KS) theory possible.
The most common approximations such as the
LDA\cite{KohnSham}, GGA\cite{GGA} and hybrid schemes\cite{Becke1993}
perform generally well, but usually give very poor results for electron
correlation alone. In particular they give completely incorrect physics
for van der Waals (vdW) dispersion physics, which governs the weak bonds
between widely separated systems. This physics can be important
in systems where there is a realm of near-zero density
between sub-systems, such as in stretched molecules or lattices.
The vdW physics is reintroduced in the popular
vdW-DF\cite{Dion2004,*Rydberg2000,*Rydberg2003,*Langreth2005}
group of functionals, however such methods fail to reproduce
the correct exponent\cite{Dobson2006,*Gould2008,*Gould2009}
for vdW power laws $U=-C_pD^{-p}$ in zero-gap systems
with at least one long and one short dimension, such as
thin slab geometries and
nano-wires\cite{[See section 4 of ]Dobson2011-JPCM}.

An alternative approach to total energy calculations is to: i) solve
for a groundstate under a given scheme (e.g. LDA)
to evaluate $\VKS(\vr)$ and, via the KS Hamiltonian
$\hh=-\half\nabla^2+\VKS(\vr)$, to evaluate the orbitals
and KS energies through $\hh\psi_i=\epsilon_i\psi_i$
and the density through $n(\vr)=\sum_if_i|\psi_i(\vr)|^2$
where $f_i$ is the occupation number of orbital $i$;
ii) recalculate the energy using the so-called exact exchange (EXX)
functional for exchange and a different functional for correlation.
Here we use the Hartree and exchange pair density
$n_{2\rm{Hx}}(\vr,\vrp)=n(\vr)n(\vrp)-|\sum_if_i\psi_i(\vr)\psi_i(\vrp)|^2$
to define the energy terms
$E^{\rm{H}}+\Ex=\half\int\frac{\d\vr\d\vrp}{|\vr-\vrp|}
n_{2\rm{Hx}}(\vr,\vrp)$
and we set the EXX total energy to
$E^{\EXX}=\int\d\vr [-\half \sum_i\psi_i(\vr)\nabla^2\psi_i(\vr)
+n(\vr)\VKS(\vr)] + E^{\rm{H}}+\Ex$.

Thus the total energy of a given
system can be calculated exactly from the KS potential, with the
exception of one term: the correlation energy, defined here through
$\Ec=E-E^{\EXX}$ where $E$ is the true groundstate energy of the system.
The correlation energy term essentially bundles the
``difficult'' physics of the true many-electron system
into a single term, which is a highly non-local
functional of the density and/or Kohn-Sham orbital
wavefunctions, and must be approximated.
An \emph{ab initio} way to evaluate correlation
energies is to use time-dependent DFT\cite{RungeGross} via the linear
density-response function, the fluctuation-dissipation theorem, and the
adiabatic connection formula to form the ``ACFD'' functional.
In recent years there has been a large increase in the
use of ACFD functionals, particularly for the evaluation
of vdW dispersion. The majority of these also make use of the
direct random-phase approximation (dRPA) which we define later.
A good discussion on, and summary of the ACFD-dRPA approach can be
found in \rcite{Eshuis2012}, although initial calculations on
inhomogeneous systems were carried out more than a
decade ago\cite{Dobson1999,*Furche2001,*Miyake2002}.

Theoretically exact applications of the ACFD involve
the unknown dynamic exchange-correlation kernel $\fxc$,
a two point \emph{function} defined as the the second
functional derivative of the xc energy via
$\fxc(\vr,\vrp;t-t')=\delta^2\Exc/[\delta n(\vr,t)\delta n(\vrp,t')]$.
The concept of the xc kernel can also be extended to
current-response theory where the tensor kernel $\tFxc$
is known\cite{VignaleKohn,*Vignale1997} to be a more
`amenable' functional of the density.
In practice $\fxc$ must be approximated, and the commonly employed
dRPA involves setting $\fxc\aeq 0$.
Perhaps surprisingly, the ACFD-dRPA functional has generally
performed well for calculating energy differences,
but not so well for absolute energies.
Through the years various approximations have been proposed
for the $\fxc$ kernel, including the ALDA\cite{ALDA},
energy-optimised kernel\cite{Dobson2000} and the
Petersilka, Gossman and Gross exchange kernel\cite{PGG}.
These have met with varying degrees of success in different systems,
but none has worked well in a wide range of systems.
More recently the exact exchange kernel $\fxc\aeq f^{\xrm}$ has been
evaluated\cite{Hellgren2008,Hesselmann2010,Hesselmann2011,Bleiziffer2012}
in the time-dependent EXX (tdEXX) approach
leading, via the ACFD functional, to excellent results for
correlation energies of atoms and molecules.
However this kernel is very difficult [$O(N^5)/O(N^6)$ in molecular
basis function language] to calculate in practice, requiring
inversion of the response or solutions of non-linear eigen-equations.
Similarly, alternative approaches such as RPAx\cite{RPAx}
and SOSEX\cite{Gruneis2009} exist to improve on the ACFD-dRPA by
including many-electron exchange but again these are numerically
more difficult problems than the dRPA.

The ISTLS formalism\cite{ISTLS,ISTLS-Current}, extending a total
energy method for jellium\cite{STLS} to general systems,
was developed as a means of approximating the dynamic interactions
in a sophisticated manner by making use of a self-consistent
pair-correlation function. As shown in \rcite{ISTLS-Current}
it is equivalent to self-consistently approximating
$\tFxc$ in an ACFD functional and it has so far
enjoyed success in semi-homogeneous test
systems\cite{ISTLS,Constantin2008Surface,%
Constantin2008Dimension,Constantin2011}, most notably correctly
reproducing the difficult transition from a three-
to a two-dimensional metal, something the dRPA fails to do.
In some sense the ISTLS represents
the `next step' of ACFD-like approximation: introducing
self-consistent physics to the \emph{dynamic} tdDFT calculation
in a rigorous manner through $\tFxc$, rather than
deriving $\fxc$ or $\tFxc$ from the groundstate calculation.

In the original paper\cite{ISTLS} on the method,
the ISTLS functional was also tested on the helium atom
where it performed very well, calculating the correlation
energy to within 0.1mHa.
Advances in computing power and improvements in numerical
techniques have since allowed for wider testing. Here
we discuss the implementation of the functional in
spherical systems, and test it in a set of spherically symmetric
neutral atoms and ions, including spin-polarised systems such as
atomic sodium and lithium.

The Kohn-Sham equations $\hh\psi_i=\epsilon_i\psi_i$
can be used to generate the one-electron like orbitals of a
system with a time-invariant KS potential $\VKS$.
In the absence of a magnetic field but the
presence of a small perurbation to $\VKS$ of form
$\delta V(\vr,t)=\delta V(\vr) e^{i\omega t}$ we can write
the change in density of the system as $\delta n=\int\d\vrp
\chi_0(\vr,\vrp;\omega)\delta V(\vrp)
\equiv\int\d\vrp \vnu_0(\vr,\vrp;\omega)\cdot\vnabla\delta V(\vrp)$
where $\chi_0$ is the bare (non-interacting) density-density response
of the system, and $\vnu_0$ is the bare vector response.
The change in current can be defined via $\delta\vj
=i\omega\int\d\vrp \tP_0(\vr,\vrp;\omega)\vnabla\delta V(\vrp)$
where $\tP_0$ is the bare current-current response.
Using tensor notation\footnote{Here and henceforth
we define indices $\mu,\nu\in(x,y,z)$, vectors to be bold $\vec{v}$,
and tensors to be upright sans-serif $\ten{T}$.
The tensor $\ten{T}=\vec{v}\otimes\vec{u}$ has elements
$T_{\mu\nu}=v_{\mu}u_{\nu}$, the vector $\vec{u}=\vec{v}\cdot\ten{T}$
has elements $u_{\mu}=v_{\nu}T_{\nu\mu}$ and
$\ten{A}:\ten{B}=\sum_{\mu\nu}A_{\mu\nu}B_{\nu\mu}$ is scalar.},
it follows from these expressions that
$\chi_0=-\vnabla'\cdot\vnu_0$ and $\vnu_0=-\vnabla\cdot\tP$.
Each of these has an interacting equivalent eg. $\chi_{\lambda}$
which corresponds to the response a related system with
electron-electron Coulomb interactions of strength $\lambda$
but with the groundstate density unchanged. When $\lambda=1$ these are
equivalent to the response of the system to a change in the
\emph{external} potential.

The ACFD correlation functional can be defined as
\begin{align}
\Ec=&\int_0^1\d\lambda \int_0^{\infty}\frac{\d s}{\pi} \int\d\vr\d\vrp
\Phi_{\lambda}(\vr,\vrp,is)
\end{align}
with integrand\cite{Note1}
$\Phi_{\lambda}(\vr,\vrp,\omega)
=[\chi_{\lambda}-\chi_0](\vr,\vrp;\omega) v(|\vrp-\vr|)
\equiv[\vnu_{\lambda}-\vnu_0](\vr,\vrp;\omega) \cdot \vnabla' v(|\vrp-\vr|)
\equiv[\tP_{\lambda}-\tP_0](\vr,\vrp;\omega):\tV(|\vrp-\vr|).$
Here $v(R)=1/R$ is the Coulomb potential and
$\tV(R)=-\vnabla\otimes\vnabla v(R)$ is its tensor equivalent.
We can explicitly write the bare density-density and density-current
reponses as
\begin{align}
\chi_0(\vr,\vrp;is)=&2\Re\sum_i f_i
\psi_i^*(\vr)\psi_i(\vrp)G_i(\vr,\vrp),
\label{eqn:chi0Def}
\\
\vnu_0(\vr,\vrp;is)=&\Im\sum_i f_i \big[
\psi_i^*(\vr)\psi_i(\vrp)\vnabla'G_i(\vr,\vrp)
\nonumber\\&
-G_i(\vr,\vrp)\vnabla'\psi_i^*(\vr)\psi_i(\vrp)
\big]/s
\label{eqn:nu0Def}
\end{align}
where $G_i$ is short-hand for the bare one-electron Greens function
$G(\vr,\vrp;\epsilon_i+is)$, a solution of
$[\hh - \Omega]G(\vr,\vrp;\Omega)=\delta(\vr-\vrp)$.
The current-current response $\tP_0$ has a similar expression.
The interacting responses are defined via
\begin{align}
\chi_{\lambda}=&\chi_0 + \chi_0\star(\lambda v + \fxc_{\lambda})\star\chi_{\lambda}
\label{eqn:chilambda}
\\
\tP_{\lambda}=&\tP_0 + \tP_0\star(\lambda\tV+\tFxc_{\lambda})\star\tP_{\lambda}
\label{eqn:Plambda}
\end{align}
where $A\star B\equiv\int\d\vx A(\vr,\vx)B(\vx,\vrp)$
and we take tensor products where appropriate. It is only in
this relationship between the interacting and non-interacting
case that the xc kernel is involved.

The ISTLS scheme can be written\cite{ISTLS-Current}
as a tensor $\tFxc$ of form
\begin{align}
\lambda\tV+\tFxc_{\lambda}=&\frac{1}{s^2}g_{\lambda}(\vr,\vrp)
\vnabla \frac{\lambda}{|\vr-\vrp|}\otimes \vnabla',
\label{eqn:kernel}
\\
g_{\lambda}(\vr,\vrp)=&n_{2\lambda}(\vr,\vrp)/[n(\vr)n(\vrp)]
\label{eqn:gl}
\end{align}
where $n_{2\lambda}$ is the interacting groundstate pair density at
coupling strength $\lambda$ and $n$ is the groundstate density.
Here we self-consistently calculate the dynamic interactions
via the pair density $n_{2\lambda}$ calculated by
the fluctuation-dissipation theorem
\begin{align}
n_{2\lambda}(\vr,\vrp)=&n(\vr)n(\vrp)-\delta(\vr-\vrp)\n(\vr)
\nonumber\\&
-\int\frac{\d s}{\pi} \chi_{\lambda}(\vr,\vrp;is),
\label{eqn:n2}
\\
\chi_{\lambda}(\vr,\vrp;is)=&(\nabla\otimes\nabla'):\tP_{\lambda}(\vr,\vrp).
\label{eqn:chiP}
\end{align}
In practice we must iterate these equations: i) set $g_{\lambda}\aeq g_0$
(ie. Hartree and exchange only) such that
$g_0(\vr,\vrp)=1-[\n(\vr)\n(\vrp)]^{-1}
|\sum_if_i\psi_i(\vr)\psi_i^*(\vrp)|^2$, ii) calculate $\tP_{\lambda}$
via \eqref{eqn:Plambda} and \eqref{eqn:kernel},
iii) use $\tP_{\lambda}$ to calculate a new $g_{\lambda}$
via \eqref{eqn:n2} and iv) use the new $g_{\lambda}$ in ii)
and repeat until convergence is reached.

Making use of \eqref{eqn:kernel} and \eqref{eqn:chiP}
we can transform \eqref{eqn:Plambda} into
$\chi_{\lambda}=\chi_0+Q_{\lambda}\star\chi_{\lambda}$
where $Q_{\lambda}(\vr,\vrp)=\int\d\vx \vnu_0(\vr,\vx)\cdot
\vF_{\lambda}(\vx,\vrp)$ and
\begin{align}
\vF_{\lambda}(\vr,\vrp)=&g_{\lambda}(\vr,\vrp)
\vnabla\frac{\lambda}{|\vr-\vrp|}.
\label{eqn:Fl}
\end{align}
Thus it is possible to evaluate the ISTLS equations
using only $\chi_0$ and $\vnu_0$
and not the full tensor current-current response $\tP_0$.
This form of the equations is the original\cite{ISTLS}
approach to ISTLS calculations.
It should be noted that
the Petersilka-Gossman-Gross (PGG) kernel\cite{PGG} can be defined
in a similar manner with
$Q_{\lambda}(\vr,\vrp)=\int\d\vx \vnu_0(\vr,\vx)\cdot
\vnabla_x g_{0}(\vx,\vrp)\frac{\lambda}{|\vx-\vrp|}
\equiv\int\d\vr\chi_0(\vr,\vx)\frac{\lambda g_{0}(\vx,\vrp)}{|\vx-\vrp|}$.

In spherically symmetric atoms we can separate the orbitals as
$\psi_i(\vr)\equiv\psi_{nlm}(\vr)=R_{nl}(r)Y_{lm}(\vrh)$
and $\epsilon_i\equiv\epsilon_{nl}$
where $Y_{lm}$ is a spherical harmonic function.
The potential is $\VKS(\vr)\equiv\VKS(r)$ and the radial function
satisfies $\hh_l R_{nl}(r)=\epsilon_{nl} R_{nl}(r)$ where
$\hh_l \equiv -\half \{ r^{-1}\Dp{r}\Dp{r}r-l(l+1)r^{-2}\} + \VKS(r)$
and $\Dp{r}\equiv \partial/\partial r$.
It follows from the properties of spherical harmonics and
the definition of the Greens function that
$\sum_m\psi_{nlm}^*(\vr)\psi_{nlm}(\vrp)=
\frac{2l+1}{4\pi}P_l(x)\gamma_{nl}(r,\rp)$
and $G(\vr,\vrp;\Omega)=\sum_l \frac{2l+1}{4\pi}P_l(x)
G_l^{\Omega}(r,\rp)$
where $x=\vrh\cdot\vrhp$,
$P_l(x)$ is a Legendre polynomial of order $l$
and we use the short-hand $\gamma_{nl}(r,\rp)=R_{nl}(r)R_{nl}(\rp)$.
Here $G_l^{\Omega}$ satisfies
$[\hh_l-\Omega]G_l^{\Omega}(r,\rp)=\delta(r-\rp)/(r\rp)$.
It also follows from the symmetry of the system that
\begin{align}
\chi_{\lambda}(\vr,\vrp;is)=&\sum_L \frac{2L+1}{4\pi}P_L(x)
\chi_{\lambda L}(r,\rp;is)
\label{eqn:chi0L}
\\
\vnu_{\lambda}(\vr,\vrp;is)=&\sum_L \frac{2L+1}{4\pi}P_L(x)
[ \nur_{\lambda L}\vrhp + \nup_{\lambda L}\vrp_{\perp} ].
\label{eqn:nu0L}
\end{align}
where $\vrp_{\perp}=\vrh-(\vrh\cdot\vrhp)\vrhp=\vrh-x\vrhp$.
Thus the response equation is diagonal in $L$
and $\chi_{\lambda L}=\chi_{0L} + Q_{\lambda L}\starr \chi_{\lambda L}$
where $A\starr B\equiv \int_0^{\infty}R^2\d R A(r,R)B(R,\rp)$.

Making use of the completeness of the polynomials $P_l(x)$
we define the bare ($\lambda=0$) responses through
\begin{align}
\chi_{0L}(r,\rp;is)=&2\sum_{nll'} K_{ll'}^L \gamma_{nl}\Re G_{l'}^{\epsilon_{nl}+is}
\\
\nur_{0L}(r,\rp;is)=& \frac{1}{s}\sum_{nll'} K_{ll'}^L\{
\gamma_{nl}[\Dp{\rp}\Im G_{l'}^{\epsilon_{nl}+is}]
\nonumber\\&
-[\Dp{\rp}\gamma_{nl}]\Im G_{l'}^{\epsilon_{nl}+is}
\}
\\
\nup_{0 L}(r,\rp;is)=&\frac{1}{s\rp}\sum_{nll'} (\beta_{l'l}^L-\beta_{ll'}^L)
\gamma_{nl}\Im G_{l'}^{\epsilon_{nl}+is}.
\end{align}
The Clebsch-Gordan-like coefficients $K_{ll'}^L$ and $\beta_{ll'}^L$ are
defined as
$K_{ll'}^L\allowbreak=\frac{(2l+1)(2l'+1)}{4\pi(2L+1)}\allowbreak
\int_{-1}^1\d x P_lP_{l'}P_L$ and
$\beta_{ll'}^L\allowbreak=\frac{(2l+1)(2l'+1)}{4\pi(2L+1)}\allowbreak
\int_{-1}^1\d x D_lP_{l'}P_L$ where
$D_l\equiv[\Dp{x}P_l(x)]$.
We can similarly expand the vector kernel \eqref{eqn:Fl}
of the ISTLS scheme as
\begin{align}
\vF(\vr,\vrp)=&\sum_{L} \frac{2L+1}{4\pi}P_L(x)
[\Fr_{L}(r,\rp)\vrh + \Fp_{L}\vr_{\perp}]
\label{eqn:FL}
\end{align}
where $\Fr_{L}=\sum_{ll'} K_{ll'}^L g_{\lambda l}[\Dp{r}v_{l'}]$
and $\Fp_{L}=\sum_{ll'} \beta_{l'l}^L g_{\lambda l}v_{l'}/r$.
We define $g_{\lambda l}$ through \eqref{eqn:gl} and
\eqref{eqn:n2} but with $\chi_{\lambda l}(r,\rp)$ only,
and use the Legendre expansion of the Coulomb potential
$1/|\vr-\vrp|=\sum_l v_l(r,\rp)\frac{(2l+1)P_l(x)}{4\pi}$ to define
$v_l=\frac{4\pi}{2l+1}\min(r,\rp)^l\max(r,\rp)^{-(l+1)}$.
Finally, using \eqref{eqn:nu0L} and \eqref{eqn:FL}
we find\cite{[Details in Chapter 7.3 of ]GouldThesis}
\begin{align}
Q_{\lambda L}=&\nur_{0 L}\starr \Fr_{L}
+ \hat{\kappa}[\nup_{0 L}\starr \Fp_{L}]
- [\hat{\kappa}\nup_{0 L}]\starr[\hat{\kappa}\Fp_{L}]
\label{eqn:QL}
\end{align}
where $\hat{\kappa}f_L\equiv K_{L1}^{L+1} f_{L+1} + K_{L1}^{L-1} f_{L-1}$.

We note that, with the exception of the self-consistency
condition [defined via \eqref{eqn:n2}], all
terms are diagonal in $s$ but couple together different $l$
and involve convolutions over radial co-ordinate $r$. This allows
us to evaluate $\chi_{\lambda L}(r,\rp;is)$ from the sets
$\{\chi_{0 l}(r,\rp;is)\}_l$ and $\{\vnu_{0 l}(r,\rp;is)\}_l$
provided the set $\{g_{\lambda l}\}_l$ is already known.
Once $Q_{\lambda L}(r,\rp;is)$ is calculated
the system is diagonal in $L$ and convolutions are only ever taken
across $r$. In spin-polarised systems we must also introduce
spin $\sigma=\uparrow\downarrow$ such that
all radial coordinates are replaced by $r\sigma$ and
convolutions include a sum over spin.

To solve such a system numerically, we choose a grid of up to 512 radial
points, and solve for the groundstate using the method of
Krieger, Li and Iafrate\cite{KLI1992} (KLI).
The KLI approximation predicts $E^{\EXX}$ quite accurately,
and reproduces the correct $-1/r$ tail in atoms, a feature not
present in LDA or GGA calculations. As such we feel it is
an ideal starting point for these calculations.

The grid $\{r_i\}$, its weights $\{w_i\}$,
the radial orbital wavefunctions $R_{nl}(r_i)$,
KS energies $\epsilon_{nl}$, and the KS potential $\VKS(r_i)$ are then
stored for later use in the calculation of $\chi_0$ and
$\vnu_0$. The Greens function can be solved quickly
at arbitrary $l$ and $\Omega$ via a shooting method such that
\begin{align}
  G_l^{\Omega}(r,\rp)=&\frac{1}{2r\rp \textrm{Wr}}\begin{cases}
    I(r)O(\rp) & r<\rp
    \\
    O(r)I(\rp) & r\geq \rp
  \end{cases}
\end{align}
where $\textrm{Wr}=I\Dp{r}O-O\Dp{r}I$
and $I(r)$ and $O(r)$ are ``inner'' or ``outer'' solutions of
$[\hh_l-\Omega]X(r)=0$
with the boundary conditions $I(r\to 0)\propto r^l$ and
$O(r\to\infty)=0$.
Its radial derivative is then
$\Dp{\rp}G_l^{\Omega}=D_l^{\Omega}-G_l^{\Omega}/\rp$ where
\begin{align}
  D_l^{\Omega}(r,\rp)=&\frac{1}{2r\rp \textrm{Wr}}\begin{cases}
    I(r)\Dp{\rp}O(\rp) & r<\rp
    \\
    O(r)\Dp{\rp}I(\rp) & r\geq\rp
  \end{cases}.
\end{align}

We choose a set of abcissae and weights for $s$ based
on a Clenshaw-Curtis quadrature scheme, chosen for
its accuracy in integrating Lorentz functions,
such that convergence is reached using at most 50 points.
We also exploit the fact that the system is diagonal
in $s$ to calculate and store response functions at a single $s$ only
and cumulatively evaluate integrals for the pair density
and correlation energy. The method is also
diagonal in $\lambda$ and we solve to high accuracy
using $\lambda=\frac{1}{3},\frac{2}{3},1$ with appropriate weights.
We must also choose a cutoff in $L$ which we set at {{$L_{\max}=6$.
In all tested cases the contribution to the energy
from the $L=5$ term is under 0.5\%, with at least 97\%
of the energy accounted for by $L\leq 3$.}}

Calculations are thus performed as follows: 1) for
each $l$ form the matrices
$\chi_{0l}(r_i,r_j;is)$, $\nur_{0l}(r_i,r_j;is)$ and
$\nup_{0l}(r_i,r_j;is)$ and, at the first iteration,
$g_{\lambda l}\aeq g_{0 l}(r_i,r_j)$;
2) take the stored response functions and
pair densities $\{g_{\lambda l}\}_l$, then use quadrature to form
$Q_{\lambda L}(r_i,r_j)$ via \eqref{eqn:QL};
3) solve the matrix equation
$\chi_{\lambda L ij}=\chi_{0 L ij} + \sum_kQ_{\lambda L ik}w_k\chi_{\lambda L kj}$,
repeating 1)-3) for each $L$ and each $s$
4) calculate new values for $\{g_{\lambda l}\}_l$ through a weighted mix
of the existing data and the newly evaluated
[via \eqref{eqn:n2}] $\{g_{\lambda l}\}_l$;
5) repeat from 1) until converged;
6) reset $\{g_{\lambda l}$ and repeat from 1) for a new $\lambda$.
Typically it takes between four
and six iterations mixing 70\% new and 30\% old
pair density to converge a correlation energy. It is worth
noting that at each stage we impose symmetry under exchange
of $r$ and $\rp$ on each $g_{\lambda l}$. While formally this may
differ slightly from the true ISTLS method, tests indicate
that the correlation energy remains virtually unchanged,
while convergence is improved.

\begin{table}[h] 
\caption{Correlation energies (in -mHa) for spherical
atoms and ions. Includes the mean absolute error \% (MAE\%)
for the neutral atoms (N), ions (I) and all atoms
and ions together. He${}^*$ is the extrapolation to $Z=\infty$
for a two-electron system.\label{tab:Ec}}
\begin{ruledtabular}
\begin{tabular}{|lrrrrr|}
Atom & RPA & PGG & ISTLS & tdEXX${}^a$ & Exact${}^b$ \\
\hline
He & 84.0 & 44.9 & 42.3 & 44 & 42.0\\
Li & 113 & 49 & 41 & - & 45\\
Be & 181 & 104 & 79 & 102 & 94\\
N & 336 & 145 & 191 & - & 188\\
Ne & 585 & 331 & 405 & 389 & 390\\
Na & 612 & 329 & 413 & - & 396\\
Mg & 672 & 374 & 458 & 445 & 438\\
P & 833 & 418 & 563 & - & 540\\
Ar & 1071 & 578 & 744 & 721 & 722\\
\hline
MAE\% N & 76 & 15 & 5 & - & \\
\hline
H${}^{-}$ & 74.9 & 43.6 & 36.4 & - & 42.0\\
Li${}^{+}$ & 86.7 & 45.3 & 42.8 & - & 43.1\\
Be${}^{2+}$ & 88.3 & 45.6 & 43.7 & - & 44.3\\
Ne${}^{8+}$ & 91.1 & 46.1 & 45.4 & - & 44.7\\
Hg${}^{78+}$ & 92.4 & 46.3 & 46.2 & - & 46.5\\
He${}^*$ & 92.7 & 46.4 & 46.4 & - & 46.9\\
Be${}^{+}$ & 124 & 51 & 37 & - & 47\\
Li${}^{-}$ & 146 & 84 & 69 & - & 73\\
B${}^{+}$ & 207 & 120 & 86 & - & 111\\
Na${}^{+}$ & 582 & 323 & 404 & - & 389\\
Mg${}^{+}$ & 623 & 331 & 422 & - & 400\\
\hline
MAE\% I & 94 & 7 & 7 & - & \\
\hline
\hline
MAE\% & 86 & 11 & 6 & - & \\
\end{tabular}
\end{ruledtabular}\\
${}^a$ From \Rcite{Hellgren2008},
${}^b$ From \Rcites{AtomCorrelation,Jiang2007,HeIsoCorrelation,Komasa2002}
\end{table}

In Table \ref{tab:Ec} we present correlation energies for a variety of
spherically symmetric systems. We compare the ISTLS energies
with those from the dRPA and PGG calculated using the same code,
with tdEXX energies from \rcite{Hellgren2008},
and with `exact' correlation energies from benchmark
methods\cite{AtomCorrelation,Jiang2007,HeIsoCorrelation,Komasa2002}.
{{We also include an extrapolation to the $Z=\infty$ case for
the Helium isolectronic series (labeled He${}^*$)
by fitting $E_c(Z)$ vs. $1/Z$ for $Z\geq 3$.}}
We have included only those atoms and ions that converged under
the ISTLS self-consistency loop with a reasonable mixing
coefficient and thus reasonable time.
For benchmarking we compared our dRPA results with those of
Jiang and Engel\cite{Jiang2007} and found agreement well within
expected methodological bounds.

In general the ISTLS does very well for correlation energies, outperforming
the dRPA in all tested systems, and the PGG in all but a few systems.
In all the systems bar He where comparable tdEXX results were
available\cite{Hellgren2008} it outperforms the ISTLS,
however this accuracy comes at much greater computational expense.
ISTLS performs less well for ions than for atoms, with the greatest error
in {{Be${}^+$ and B${}^+$}}. It is possible that, in these cases,
the ISTLS iterations
converge to an incorrect result, however testing this is difficult.
{{For C${}^{2+}$ the ISTLS method did not converge at all, most
likely due to numerical instabilities in the high-density core region.}}
It is worth noting that the ISTLS \emph{always} pulls the PGG
results back towards the true value, albeit overly so in some cases.
While the PGG approximation performs slightly better than ISTLS for
some of the smaller systems tested here, it is known to break down
in bulk systems, particularly
low density metals where it under-correlates\cite{Lein2000}.
This failure can be seen in the trend presented here,
where the relative absolute PGG error increases with system size
while ISTLS improves.
By contrast the ISTLS performs consistently well for jellium\cite{STLS},
metallic surface energies\cite{Constantin2008Surface}, across
two- and three-dimensional metals\cite{Constantin2008Dimension}, and here
in the spherical atoms and ions.

The numerical cost of the ISTLS functional scales with system
size in a similar manner to standard ACFD-dRPA methods,
but with a larger pre-factor and slightly larger memory requirements.
In the best case scenario, the ISTLS can scale as $O(N^4)$,
while tdEXX and RPAx can scale as $O(N^5)$, a saving of one order.
Our ISTLS calculations took between ten and twenty times as long as
the ACFD-dRPA and used around five times the memory.
The detailed method presented here may point the way to implementation
in more general geometries involving expansions on
Gaussian-type and Slater-type orbitals\cite{SCMOM-1,*SCMOM-3}.
Implementation in existing plane-wave based bulk ACFD-dRPA codes
should also be possible, albeit with non-trivial changes.

Overall, we believe that the ISTLS is an excellent candidate for
a `next step' functional, going beyond the dRPA.
The tests on spherical systems further confirm
its versatility, showing accurate results in systems with
vastly different physics to those previously tested.
With work on efficiencies and implementation it
could, in future, provide viable benchmark calculations for
electronic systems where existing high-level
methods, such as the popular ACFD-dRPA, fail to achieve
the desired level of accuracy and where wavefunction methods are
too difficult.

The authors were supported by ARC Discovery Grant DP1096240.

\bibliography{vanDerWaals,ACFDT,DFT,Wannier,Misc,Experiment,Hybrid,%
QMGeneral,ISTLS}

\end{document}